\documentclass[final]{IEEEtran}

\newlength \figwidth
\usepackage{cite}
\ifCLASSINFOpdf
  \usepackage[pdftex]{graphicx}
	\usepackage{epstopdf}
  \graphicspath{{../Figures/}}
  \DeclareGraphicsExtensions{.eps}
\else
  \usepackage[dvips]{graphicx}
\fi
\usepackage[cmex10]{amsmath}
\usepackage{amssymb}
\usepackage{amsthm}
\usepackage{url}
\usepackage{dsfont}
\usepackage{tabulary}
\usepackage{multirow}
\usepackage{color}

\usepackage{color}
\usepackage{times}
\usepackage{verbatim}
\usepackage{floatflt}
\usepackage{enumerate}
\usepackage{array}
\usepackage{multicol,afterpage,wrapfig} 
\usepackage{tikz}
\usepackage{algorithm}
\usepackage[noend]{algpseudocode}
\makeatletter
\def\BState{\State\hskip-\ALG@thistlm}
\makeatother
\usepackage{amsmath,amssymb,eucal}
\usepackage[utf8]{inputenc}
\usepackage{epsfig}
\usepackage{exscale}
\usepackage{latexsym}
\usepackage{verbatim}
\usepackage{amsfonts}
\usepackage{multirow}
\usepackage{cite}
\usepackage{balance}
\usepackage{color}
\usepackage{transparent}
\usepackage{import}
\usepackage{mathtools}
\usepackage{tikz}
\usetikzlibrary{automata,arrows,positioning,calc}
\usepackage{bbm}
\usepackage[printonlyused,withpage]{acronym}
\usepackage{tabu}
\usepackage{longtable}
\usepackage{balance}
\usepackage{footnote}
\usepackage{tablefootnote}
\usepackage[font=footnotesize]{caption}
\usepackage{enumitem}

\def\BibTeX{{\rm B\kern-.05em{\sc i\kern-.025em b}\kern-.08em
    T\kern-.1667em\lower.7ex\hbox{E}\kern-.125emX}}
    
\renewcommand{\IEEEQED}{\IEEEQEDopen}

\newcommand*\xbar[1]{%
  \hbox{%
    \vbox{%
      \hrule height 0.5pt 
      \kern0.36ex
      \hbox{%
        \kern-0.12em
        \ensuremath{#1}%
        \kern-0.12em
      }%
    }%
  }%
}

\newfont{\bbb}{msbm10 scaled 500}

\newfont{\bb}{msbm10 scaled 1100}

\newcommand{\executeiffilenewer}[3]{%
\ifnum\pdfstrcmp{\pdffilemoddate{#1}}%
{\pdffilemoddate{#2}}>0%
{\immediate\write18{#3}}\fi%
}

\usetikzlibrary{arrows,calc}
\allowdisplaybreaks

\IEEEoverridecommandlockouts
\begin{document}
\pagenumbering{gobble}

\newtheorem{Theorem}{\bf Theorem}
\newtheorem{Corollary}{\bf Corollary}
\newtheorem{Remark}{\bf Remark}
\newtheorem{Lemma}{\bf Lemma}
\newtheorem{Proposition}{\bf Proposition}
\newtheorem{Assumption}{\bf Assumption}
\newtheorem{Definition}{\bf Definition}
\title{Massive MIMO Unlicensed \\ for High-Performance Indoor Networks}
\author{\IEEEauthorblockN{{Adrian~Garcia-Rodriguez$^{\dag}$, Giovanni~Geraci$^{\dag}$, David~L\'{o}pez-P\'{e}rez$^{\dag}$, \\ Lorenzo~Galati~Giordano$^{\dag}$, Ming~Ding$^{\ddag}$, and Holger~Claussen$^{\dag}$}}\\
\normalsize\IEEEauthorblockA{\emph{$^{\dag}$Nokia Bell Laboratories, Ireland}} \\
\normalsize\IEEEauthorblockA{\emph{$^{\ddag}$Data 61, Australia}}}
\maketitle
\thispagestyle{empty}

\IEEEpeerreviewmaketitle
\begin{abstract}
We propose massive MIMO unlicensed (mMIMO-U) as a high-capacity solution for future indoor wireless networks operating in the unlicensed spectrum. Building upon massive MIMO (mMIMO), mMIMO-U incorporates additional key features, such as the capability of placing accurate radiation nulls towards coexisting nodes during the channel access and data transmission phases. We demonstrate the spectrum reuse and data rate improvements attained by mMIMO-U by comparing three practical deployments: single-antenna Wi-Fi, where an indoor operator deploys three single-antenna Wi-Fi access points (APs), and two other scenarios where the central AP is replaced by either a mMIMO AP or the proposed mMIMO-U AP. We show that upgrading the central AP with mMIMO-U provides increased channel access opportunities for all of them. Moreover, mMIMO-U achieves four-fold and seven-fold gains in median throughput when compared to traditional mMIMO and single-antenna setups, respectively.
\end{abstract}
\vspace*{-0.22cm}
\section{Introduction}

Factories and enterprises in all fields are heading towards the so-called \emph{Industry 4.0}, a digital transformation aiming at more efficient, automated, and flexible processes, which will in turn enable better services and increased productivity \cite{Grier2017,HavZim2017}. A key requirement of this fourth industrial revolution is the interoperability and reliable exchange of information between smart sensors, controllers, actuators, and mobile devices, made possible by a ubiquitous high-performance wireless connectivity \cite{PorHep2014,WolSauJas2017}. Due to the cost of purchasing licensed spectrum as well as data confidentiality concerns, private and public institutions alike might deem it strategically important to guarantee such in-building, site-wide fast connectivity by leveraging the unlicensed spectrum and independently running their own networks \cite{GarGerGal2017}.

Wireless communications in unlicensed bands have recently experienced a major evolution, with the appearance of new technologies alongside the omnipresent 802.11 (Wi-Fi) \cite{PerSta2013}, such as LTE-unlicensed (LTE-U) \cite{ZhaWanCai2015}, licensed-assisted access (LAA) \cite{3GPP36889}, and MulteFire \cite{MulteFireTechnicalPaper} -- where the latter also allows stand-alone unlicensed operations without a licensed carrier anchor. A common denominator in all these new technologies is the need to comply with strict unlicensed channel access regulations, which depending on the geographical region entail interleaving inefficient idle communication intervals, or performing a clear channel assessment (CCA) before transmission \cite{3GPP-RP-140808}. The latter procedure, also known as listen-before-talk (LBT), restrains the number of nodes simultaneously transmitting in a given coverage area, posing a severe limitation on the throughput and delay attainable in dense indoor scenarios.

Precisely to maximize unlicensed spectrum reuse and data rates, we propose massive MIMO unlicensed (mMIMO-U) as a solution for future wireless indoor networks. Based on massive MIMO (mMIMO), which advocates equipping access points (APs) with a large number of antennas \cite{Mar:10}, mMIMO-U also integrates new fundamental features that may be adopted by LAA, MulteFire, and Wi-Fi \cite{GerGarLop2016,GerGarLop2016ICC}. These include the capability of placing accurate radiation nulls towards coexisting nodes (i) during the CCA phase, which guarantees channel access as long as they are well placed, and (ii) during data transmission, which ensures that no interference is generated towards coexisting nodes. Our proposal directly targets the deployment of high-performance indoor networks, including those demanded by the \emph{Industry 4.0}.

While mMIMO-U is a versatile solution, capable of complementing various existing technologies, in this paper, we study its performance when paired with an indoor Wi-Fi network. In particular, we consider the following scenarios:
\begin{enumerate}[label=\Alph*.] 
\item \textit{Single-antenna Wi-Fi}, where an indoor operator deploys three single-antenna Wi-Fi APs.
\item \textit{mMIMO Wi-Fi}, where such operator replaces one of the Wi-Fi APs, the central one, with a mMIMO \mbox{Wi-Fi AP.}
\item \textit{mMIMO-U Wi-Fi}, where the central AP is upgraded to the proposed mMIMO-U Wi-Fi.
\end{enumerate}
Our study concludes that equipping just one of the APs with the proposed mMIMO-U technology substantially boosts the user throughput when compared with the other scenarios.
\vspace*{-0.3cm}
\section{System Setup}

\subsection{Deployment}

\begin{figure}[!t]
    \centering
        \includegraphics[width=0.8\columnwidth]{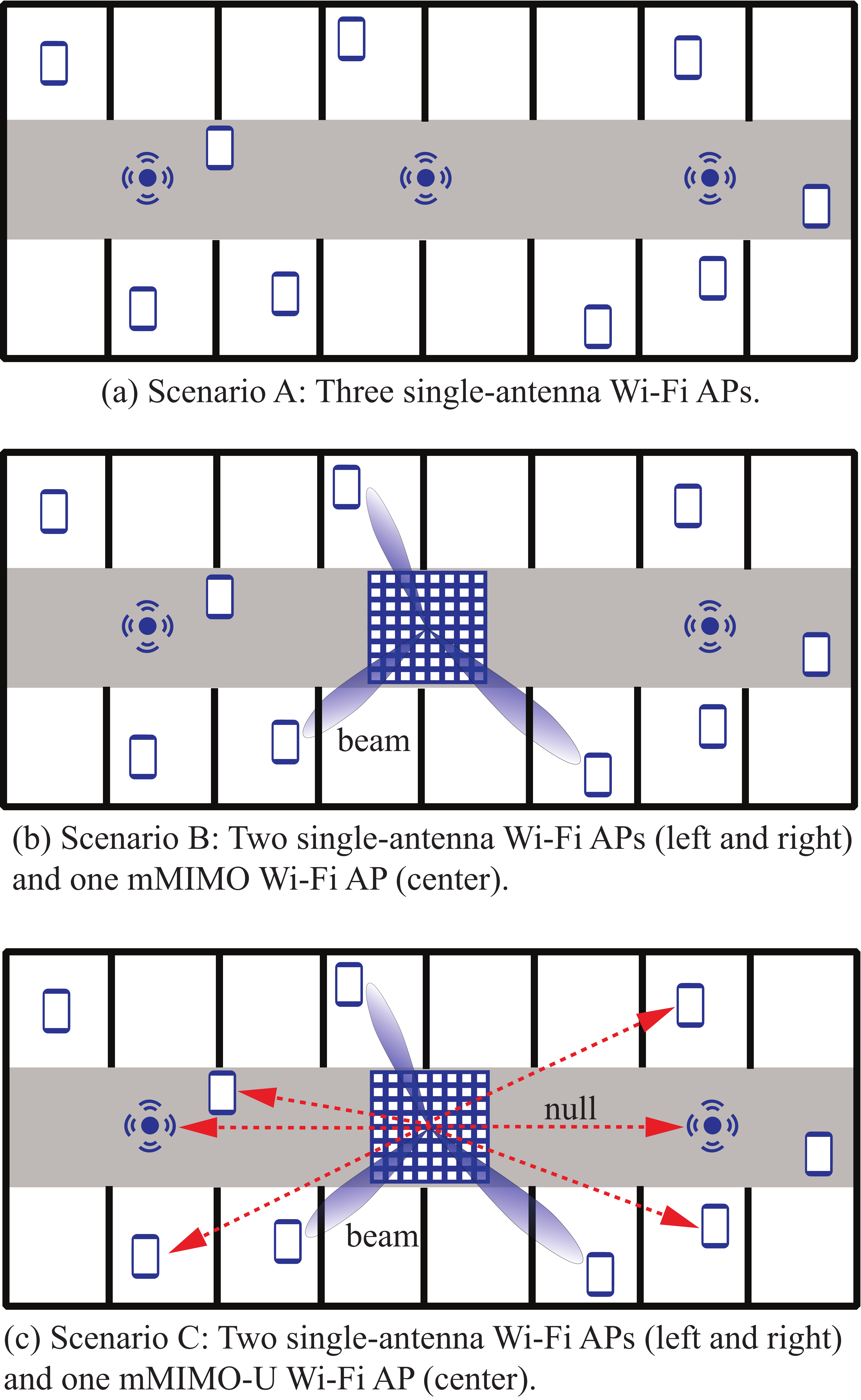}
    \caption{The three deployment scenarios considered.}
		\label{fig:SystemModel}
		\vspace*{-0.25cm}
\end{figure}

We consider the single-floor $120~\textrm{m}\times 50~\textrm{m}$ indoor hotspot network depicted in Fig.~\ref{fig:SystemModel} and operating in an unlicensed band. In this setting, which is conventionally recommended for indoor coexistence studies, an operator deploys three Wi-Fi APs on the ceiling of the central corridor to guarantee a full coverage, i.e., a minimum received signal strength (RSS) of -82 dBm for all users, also referred to as stations (STAs), located across the floor \cite{3GPP36889}. The three images in Fig.~\ref{fig:SystemModel} represent the various scenarios considered in this paper, namely: (a) single-antenna Wi-Fi, (b) mMIMO Wi-Fi, and (c) mMIMO-U Wi-Fi. We denote by $\mathcal{A}$ and $\mathcal{U}$ the sets of APs and STAs, respectively, and assume that all STAs are equipped with a single antenna. We consider that each STA has traffic available with a certain probability $P_{\textrm{tr}}$, which makes it eligible for communication. A fraction of such data traffic is to be received in downlink (DL) from the serving AP when scheduled, whereas the remaining fraction is to be transmitted in uplink (UL) towards the serving AP when the channel is available. Let $\mathcal{A}^{\star}$ and $\mathcal{U}^{\star}$ be the sets of APs and STAs transmitting signals at a given symbol interval, respectively.
STAs are served by the AP that provides the largest average RSS, and the set of STAs served by AP $a$ is denoted by $\mathcal{U}_a$. 

\vspace*{-0.1cm}
\subsection{Channel Model}

The considered indoor setup constitutes a challenging scenario for spectrum sharing due to the physical proximity between nodes. In fact, the probability of line-of-sight (LOS)  $P_{\mathrm{LOS}}$ as a function of the 3D distance $d$ in meters between any two nodes follows \cite{3GPP36889}
\begin{equation}
  P_{\mathrm{LOS}} = \left\{ 
  \begin{array}{c c l}
  1 & \quad \textrm{if} \enspace d \leq 18\\
	e^{-\frac{d-18}{27}} & \quad \textrm{if} \enspace 18 < d \leq 37\\
  0.5 & \quad \textrm{if} \enspace d > 37.\\
	\end{array} \right.
\label{eqn:PLOS}
\end{equation}
The above entails that the mutual interference between nodes reusing the same spectrum is significantly larger than that considered in more sparse outdoor deployments \cite{GerGarLop2016ICC,GerGarLop2016,GarGerGal2017}.

All propagation channels are affected by slow channel gain (comprising antenna gain, path loss, and shadowing) and fast fading. We adopt a block-fading propagation model, and assume reciprocity since UL/DL transmissions share the same frequency band. 
Let $M_i$ be the number of antennas at a given node $i\in \mathcal{A} \cup \mathcal{U}$, and $K_i$ the number of data streams it simultaneously transmits/receives. The signal $\mathbf{z}_{i} \in \mathbb{C}^{M_{i}}$ received by node $i$ at a given symbol interval is given by
\begin{align}
\mathbf{z}_{i} = \! \! \sum_{j \in \mathcal{A}^{\star} \cup \mathcal{U}^{\star}} \!\! \sqrt{P_{j}} \, \mathbf{H}_{ij}^{\mathrm{H}} \mathbf{W}_{j} \mathbf{s}_{j} \!+\! \boldsymbol{\epsilon}_{i},
\label{eqn:zi}
\end{align}
where:
\begin{itemize}
\item $\mathbf{H}_{ij} \in \mathbb{C}^{M_{i}\times M_{j}}$ denotes the channel matrix between nodes $j$ and $i$,
\item $\mathbf{W}_{j} \in \mathbb{C}^{M_{j}\times K_{j}}$ is the normalized precoding matrix employed by node $j$, 
\item $\mathbf{s}_{j} \in \mathbb{C}^{M_{j}}$ is the unit-variance signal vector, 
\item $P_j$ denotes the average transmission power of node $j$, and
\item $\boldsymbol{\epsilon}_{i} \in \mathbb{C}^{M_{i}}$ is zero-mean complex Gaussian thermal noise with variance $\sigma^2_{\epsilon}$.
\end{itemize}
We note that when both $i$ and $j$ are single-antenna nodes, all above variables reduce to scalars, and that $\mathbf{W}_{j} = 1, \forall j \in \mathcal{U}^{\star}$. For each AP $a \in \mathcal{A}$, an estimate of the channel $\mathbf{H}_{ia}$ to/from each associated STA $i \in \mathcal{U}_a$ can be obtained via pilot signals transmitted during a training phase. In this paper, such estimate is assumed to be perfect. Studies considering pilot contamination and other impairments will be part of our future work. The SINR per subcarrier of user $i$ served by the $a$-th AP can be expressed as
\begin{equation}
\text{SINR}_{i} = \frac{P_{a} \vert \mathbf{h}_{ia}^{\mathrm{H}} \mathbf{w}_{ia} \vert^{2}}{\sum\limits_{j \in \mathcal{A}^{\star} \cup \mathcal{U}^{\star} \setminus a }\!\! \hspace{-0.3cm} P_{j} \, \vert \mathbf{H}_{ij}^{\mathrm{H}} \mathbf{W}_{j} \vert^{2} + \sum\limits_{k \in \mathcal{K}_{a} \setminus i} \hspace{-0.2cm}  P_{a} \vert \mathbf{h}_{ia}^{\mathrm{H}} \mathbf{w}_{ka} \vert^{2} + \sigma^{2}_{\epsilon}},
\label{eq:SINR}
\end{equation}
with $\mathbf{w}_{ai} \in \mathbb{C}^{M_a}$ denoting the precoding vector for user $i$, and $\mathcal{K}_{a}$ the set of users scheduled by the $a$-th AP.
\section{Channel Access, Data Transmission, \\ and User Scheduling}

We now detail the channel access, transmission, and user scheduling mechanisms for each of the scenarios depicted in Fig.~\ref{fig:SystemModel}. We concentrate on describing the DL operations for brevity, since similar procedures are followed for the UL.

\subsection{Scenario A: Single-antenna Wi-Fi}

In Scenario A, illustrated in Fig.~\ref{fig:SystemModel}(a), the operator deploys three single-antenna Wi-Fi APs.

\subsubsection{Channel Access}
In order to comply with the regulations in the unlicensed band, all APs and STAs must perform LBT before transmission~\cite{3GPP-RP-140808}. In particular, a transmission opportunity is gained by node $i$ if the sum power received from all active nodes satisfies
\begin{equation}
\| \mathbf{z}_i\|^2 < \gamma_{\mathrm{LBT}},
\label{eqn:LBT}
\end{equation}
for a designated time interval $T$, where $\gamma_{\mathrm{LBT}}$ is a regulatory threshold, and $T$ is given by a distributed inter-frame space (DIFS) interval plus a random number of backoff time slots~\cite{PerSta2013}. The process in (\ref{eqn:LBT}), known as energy detection, allows for the transmission of a single node within a certain coverage area, thus limiting spatial reuse.\footnote{Additionally, a node refrains from transmission when it detects a packet preamble from any other node, where detection requires a lower power threshold $\gamma_{\textrm{preamble}}$ with a minimum SINR of -0.8 dB \cite{PerSta2013}. We account for the preamble detection mechanism in all scenarios considered in this paper.}
 
\subsubsection{Transmission}

Upon gaining channel access, each AP $a$ schedules one of its active STAs for DL transmission, and it employs a matched-filter precoder, i.e., $\mathbf{W}_{a} = \mathbf{H}_{ia}$.

\subsubsection{Scheduling}

All APs employ a round robin scheduler in the DL, selecting an active STA at random for transmission.

\subsection{Scenario B: mMIMO Wi-Fi}

In Scenario B, illustrated in Fig.~\ref{fig:SystemModel}(b), the features of the left and right APs remain unchanged, whereas the central single-antenna AP is replaced by a mMIMO AP, denoted as $x$, and equipped with $M_x$ antennas.

\subsubsection{Channel Access}

All three APs, including the central one, and all STAs employ LBT and preamble detection mechanisms in the same fashion as described in Scenario A.

\subsubsection{Transmission}

Left and right APs operate as described in Scenario A. The central mMIMO AP selects a maximum of $K_x$ active STAs for DL transmissions, and simultaneously serves them via spatial multiplexing by employing a zero-forcing (ZF) precoder. Let
\begin{equation}
\overline{\mathbf{H}}_x \triangleq \left[ {\mathbf{H}}_{1 x},\ldots,{\mathbf{H}}_{K_x x} \right],	
\label{eqn:Hbar}
\end{equation}
be the aggregate channel matrix of all STAs scheduled by the central mMIMO AP $x$. Then, the precoder $\mathbf{W}_x$ is given by
\begin{align}
\mathbf{W}_{x} = \frac{1}{\sqrt{\zeta}} \overline{\mathbf{H}}_x \left( \overline{\mathbf{H}}_x^{\mathrm{H}} \overline{\mathbf{H}}_x \right)^{-1},
\label{eqn:ZF}
\end{align}
where the constant $\zeta$  is chosen to normalize the average transmit power such that $\Vert \mathbf{W}_{x}\Vert^2 = 1$. When operating in the unlicensed spectrum, the maximum transmit power is strictly regulated and must account for the number of spatial degrees of freedom (d.o.f.) used to provide beamforming gain~\cite{FCC2013}. The central mMIMO AP therefore abides the regulations by reducing the radiated power according to the beamforming gain provided to each STA~\cite{FCC1430}, yielding a total of
\begin{equation}
P_x = P^{\mathrm{max}}_x - 10\log_{10}(N_x/K_x)~\textrm{dBm},
\label{eq:txPowermMIMO}
\end{equation}
where $P^{\mathrm{max}}_x$ is the maximum transmission power of node $x$.

\subsubsection{Scheduling}

Left, central, and right APs employ a round robin scheduler, respectively selecting a maximum of one, $K_x$, and one active STA at random for DL transmission. Instead, only one STA per cell is scheduled for UL independently of the number of antennas implemented at the Wi-Fi APs \cite{gast2005802}.

\subsection{Scenario C: mMIMO-U Wi-Fi}

In Scenario C, illustrated in Fig.~\ref{fig:SystemModel}(c), the central AP $x$ is upgraded to a mMIMO-U AP, equipped with $M_x$ antennas, and capable not only of the operations described in Section~III-B, but also of performing interference suppression both during the channel access and transmission phases \cite{GerGarLop2016}.

\subsubsection{Channel Access}

Left and right APs, as well as all STAs, perform LBT as described for Scenarios A and B. On the other hand, the central mMIMO-U AP is able to exploit its large number of transmit antennas to enhance coexistence with neighboring nodes, so that they can simultaneously reuse the spectrum. The additional capabilities at the mMIMO-U AP in terms of channel access can be outlined as: (i) blind channel covariance estimation, and (ii) enhanced LBT (eLBT).

Channel covariance estimation allows the central AP to learn the channel subspace occupied by nodes from other cells while remaining silent. Let us denote by $\mathbf{Z}_x \in \mathbb{C}^{M_x \times M_x}$ the covariance matrix of the received signal $\mathbf{z}_x$, defined as
\begin{equation}
\mathbf{Z}_x =  \mathbb{E} \left[ \mathbf{z}_x \mathbf{z}_x^{\mathrm{H}}\right],
\label{eqn:Z}
\end{equation}
where the expectation is taken with respect to the noise vector $\boldsymbol{\epsilon}_x$ and to all symbols $\mathbf{s}_j$ in (\ref{eqn:zi}). Assuming a perfect knowledge of $\mathbf{Z}_x$\footnote{In practice, an estimate of $\mathbf{Z}_x$ can be obtained via a simple average over multiple symbol intervals \cite{HoyHosTen2014}. See \cite{GerGarLop2016} for a discussion on the effect of an imperfect covariance estimation.}, the central mMIMO-U AP applies a spectral decomposition on $\mathbf{Z}_x$, obtaining its eigenvalues sorted in decreasing order ${\nu}_{i}$, $i=1,\ldots,M_x$, and its corresponding eigenvectors ${\mathbf{u}}_{i}$, $i=1,\ldots,M_x$.
The LBT phase at the central mMIMO-U AP is then enhanced (eLBT) by \emph{nulling}, i.e., filtering out, the interference received on the $N_x$ dominant directions in ${\mathbf{Z}_x}$. In other words, let
\begin{equation}
{\mathbf{\Sigma}_x} \triangleq \left[ {\mathbf{u}}_{N_x + 1},\ldots,{\mathbf{u}}_{M_x} \right],	
\label{eqn:Sigma}
\end{equation}
be the matrix whose columns contain all eigenvectors of ${\mathbf{Z}}_x$ except the $N_x$ dominant ones. With eLBT, a transmission opportunity is gained if the condition
\begin{equation}
\left\| \left( {\mathbf{\Sigma}_x} \mathbf{\Sigma}_x^{\mathrm{H}} \right) \mathbf{z}_x \right\|^2 < \gamma_{\mathrm{LBT}},
\label{eqn:eLBT}
\end{equation}
holds for a designated amount of time $T$~\cite{PerSta2013}. Provided that a sufficient number of nulls $N_x$ have been applied, the condition in (\ref{eqn:eLBT}) is met. Therefore, unlike conventional LBT operations, eLBT allows the central mMIMO-U AP to access the channel while other nodes are active.

\subsubsection*{Remark 1}

The vectors $\mathbf{u}_{i}$, $i=1,\ldots,N_x$, span the channel subspace on which the central mMIMO-U AP receives a significant power transmitted from other nodes. Due to channel reciprocity, any power transmitted by the central AP on such subspace would generate significant interference at its neighboring nodes. For this reason, whenever pursuing spectrum reuse, i.e., attempting channel access via the proposed eLBT, the central AP must also suppress the interference generated on the directions $\mathbf{u}_{i}$, $i=1,\ldots,N_x$ during data transmission. This is accomplished by sacrificing ${N_x}$ spatial d.o.f. to place radiation nulls, as detailed in Section III-C-2.

\subsubsection*{Remark 2}

\begin{figure}[!t]
    \centering
        \includegraphics[width=0.85\columnwidth]{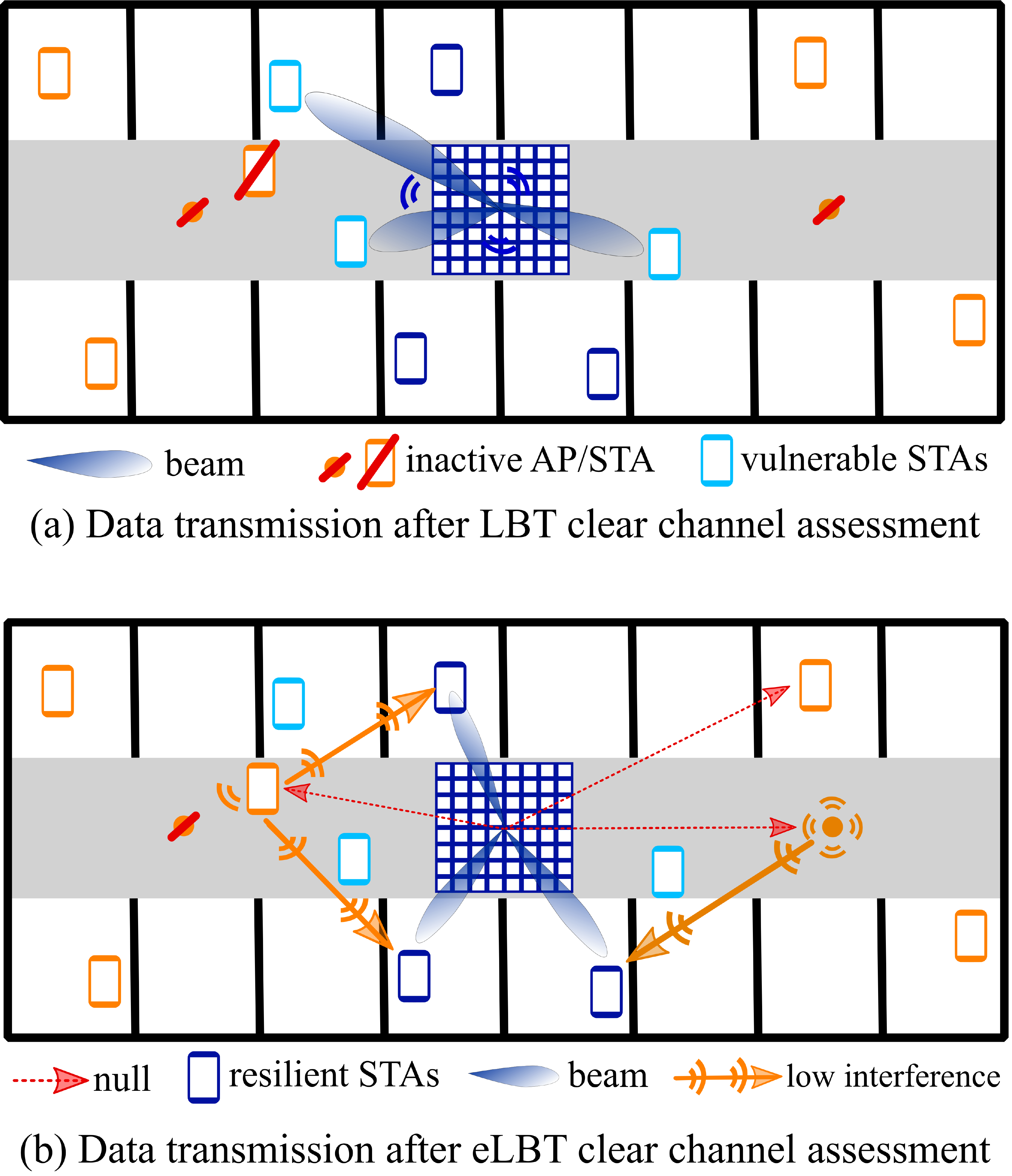}
        \caption{(a) Users that are vulnerable to strong interference are served while neighboring nodes do not transmit, i.e., after LBT. (b) Users that are resilient to some level of interference can be scheduled while neighboring nodes are active, i.e., after eLBT.}
        \label{fig:userSchedulingFig}
\end{figure}

While the proposed eLBT channel access mechanism provides the strong advantage of creating additional transmission opportunities for the central AP, this may come at the expense of increased interference at its served STAs, since the spectrum is reused while other nodes are active. This is especially true for STAs that are particularly close to nearby active nodes. In order to serve such vulnerable STAs, it is therefore advisable to periodically revert to conventional LBT channel access, based on discontinuous transmission, as illustrated in Fig.~\ref{fig:userSchedulingFig}(a) and explained in Section III-C-3.

\subsubsection{Transmission}

Left and right APs stick to single-user omnidirectional transmission/reception as in Scenario A. On the other hand, the central mMIMO-U AP (i) spatially multiplexes $K_x$ of its active STAs in DL, and (ii) employs $N_x$ of its spatial d.o.f. to suppress all interference generated on the $N_x$ dominant directions of ${\mathbf{Z}_x}$. Note that the mMIMO AP considered in Scenario B performs (i), but it does not perform (ii). Operation (ii) can be seen as forcing radiation nulls on the channel subspace spanned by $N_x$ neighboring nodes, as shown in Fig.~\ref{fig:SystemModel}(c) with the red arrows. Let 
\begin{equation}
\widetilde{\mathbf{H}}_x \triangleq \left[ {\mathbf{H}}_{1 x},\ldots,{\mathbf{H}}_{K_x x}, \mathbf{u}_{1} \ldots, \mathbf{u}_{N_x} \right],	
\label{eqn:Htilde}
\end{equation}
be a matrix containing the channels of all scheduled STAs as well as the spatial directions to null. Then, the DL precoder $\mathbf{W}_x$ at the mMIMO-U AP is given by the first $K_x$ columns of the matrix $\widetilde{\mathbf{W}}_x$ defined as
\begin{align}
\widetilde{\mathbf{W}}_x = \frac{1}{\sqrt{\zeta}} \widetilde{\mathbf{H}}_x \left( \widetilde{\mathbf{H}}_x^{\mathrm{H}} \widetilde{\mathbf{H}}_x \right)^{-1},
\label{eqn:ZFnulls}
\end{align}
where the constant $\zeta$  is chosen to normalize the average transmit power such that $\Vert \mathbf{W}_{x}\Vert^2 = 1$. As in Scenario B, the central multi-antenna AP must abide the regulations by reducing the radiated power according to the beamforming gain provided to each STA~\cite{FCC1430}, yielding a total of
\begin{equation}
P_x = P^{\mathrm{max}}_x - 10\log_{10}(M_x-N_x)/K_x~\textrm{dBm}.	
\end{equation}


\subsubsection{Scheduling}

Left and right APs employ a round robin scheduler in the downlink and, as in scenarios A and B, all cells only implement single-user MIMO UL transmissions \cite{gast2005802}. On the other hand, as explained in Remark~2, the central AP should protect those STAs particularly close to other interfering active nodes, e.g., by serving them after a conventional LBT channel access when neighboring nodes do not transmit. For this reason, in this work we introduce a transmission pattern for mMIMO-U, where the central AP alternates between LBT and eLBT channel access attempts, as illustrated in Fig.~\ref{fig:userSchedulingFig}(a) and (b), respectively. Specifically, we assume that the mMIMO-U AP accesses the channel as follows
\begin{itemize}
\item $40\%$ of the time via eLBT, serving the $40\%$ least interfered STAs with traffic available for downlink transmission. For a given AP $a$, these STAs are selected as those with the highest metric \cite{GerGarLop2016}
\begin{equation}
\beta_{i} = \frac{P_{a} \dot{h}_{i}}{\sum_{j \in \mathcal{A} \setminus a} P_{j} \dot{h}_{ji} + \sigma_{\epsilon}^2}, ~ i \in \mathcal{U}_a,
\label{eq:schedulingMetric}
\end{equation}
where $\dot{h}_{ji}$ denotes the slow fading value of the channel between nodes $j$ and $i$, and $\setminus$ denotes set subtraction. Intuitively, \eqref{eq:schedulingMetric} captures the radio distance proximity of the associated STAs to other cells \cite{GerGarLop2016}. \item $60\%$ of the time via conventional LBT operations, serving the remaining $60\%$ most vulnerable STAs.
\end{itemize}
While we have found that this choice provides an optimized performance in the considered scenario, the dynamic self-optimization of this pattern is the subject of future work. 

\vspace*{-0.3cm}
\section{Performance Evaluation}

In this section, we compare the DL performance of the three scenarios depicted in Figure~\ref{fig:SystemModel}. Specifically, we evaluate the channel access success rate for each AP, the user SINR, and the sum user throughput. A detailed list of all system parameters is provided in Table~\ref{table:parameters}. We consider two different traffic regimes, i.e., light and heavy traffic, by setting the probability of each user having data packets to transmit/receive to $P_{\text{tr}} = 0.1$ and $P_{\text{tr}} = 1$, respectively. We assume that both mMIMO and mMIMO-U APs are equipped with $M_x=36$ antennas, and schedule a maximum of four STAs for multi-user DL transmission. In practice, the allocation of d.o.f. could be dynamically optimized by trading multiplexing gain for beamforming and nulling capabilities. Moreover, we let all nodes follow the standard physical layer numerology of 802.11ac \cite{WLANStandard}.

\begin{table}
\centering
\caption{Detailed system parameters}
\vspace{-0.15cm}
\label{table:parameters}
\def\arraystretch{1.1}
\begin{tabulary}{\columnwidth}{ |p{4cm} | p{3.95cm} | }
\hline
\textbf{Parameter} 					& \textbf{Description} 		\\ \hline
\textbf{RF} 						&  							\\ \hline
AP/STA maximum TX power			& $P^{\textrm{max}} = 24/18$ dBm \cite{3GPP36889}															\\ \hline
AP and STA antenna elements         & Omnidirectional with 0 dBi \cite{3GPP36889}						\\ \hline
System bandwidth 					& 20 MHz \cite{gast2005802}															\\ \hline
Carrier frequency					& 5.18 GHz (U-NII-1) \cite{gast2005802}												\\ \hline
CCA threshold 									& $\gamma_{\textrm{LBT}} = -62$ dBm \cite{gast2005802} 				\\ \hline
Preamble detection					& $\gamma_{\textrm{preamble}}=-82$~dBm with -0.8 dB of minimum SINR \cite{gast2005802} 	\\ \hline
STA noise figure 					& 9 dB \cite{3GPP36889} 		\\ \hline
\textbf{Channel model} 				&  					\\ \hline
Path loss and probability of LOS 				& InH \cite{3GPP36843} for all links 		\\ \hline
Shadowing 										& Log-normal with $\sigma = 3/4 $ dB (LOS/NLOS)\cite{3GPP36889} 			\\ \hline
Fast fading 									& Ricean with log-normal K factor \cite{3GPP36889} and Rayleigh multipath	\\ \hline
Thermal noise 									& -174 dBm/Hz spectral density	\\ \hline
\textbf{Deployment} 									&  								\\ \hline
Floor size									& $120~\textrm{m}\times 50~\textrm{m}$					\\ \hline
AP positions								& Ceiling mounted, equally spaced in central corridor as in Fig.~\ref{fig:SystemModel}			\\ \hline
AP and STA heights								& 3 and 1.5 meters					\\ \hline
STA distribution 								& 30 uniformly deployed STAs 	\\ \hline
STA association criterion 						& Strongest received signal 	\\ \hline
DL/UL traffic fraction							& 0.8/0.2 \cite{3GPP36889}						\\ \hline
\textbf{Scenario A: single-antenna Wi-Fi} 									&  								\\ \hline
Number of antennas per AP 						& 1, 1, 1 							\\ \hline
Maximum number scheduled STAs 					& 1, 1, 1 							\\ \hline
STA scheduling 									& Round robin 				\\ \hline
\textbf{Scenario B: mMIMO Wi-Fi} 									&  								\\ \hline
Number of antennas per AP 						& 1, 36 $(6 \times 6)$, 1 							\\ \hline
Maximum number scheduled STAs 					& 1, 4, 1 							\\ \hline
Precoder 										& ZF 	\\ \hline
STA scheduling 									& Round robin \\ \hline
\textbf{Scenario C: mMIMO-U Wi-Fi} 				&  								\\ \hline
Number of antennas per AP 						& 1, 36 $(6 \times 6)$, 1 							\\ \hline
Maximum number scheduled STAs 					& 1, 4, 1 							\\ \hline
Precoder 										& ZF with interference suppression \cite{YanGerQueTSP2016}, 24 d.o.f. for nulls 	\\ \hline
STA scheduling 									& Round robin with LBT/eLBT user-based selection (Sec. III-C-3)\\ \hline
\end{tabulary}
\end{table}

\subsection{Channel Access Success Rate}

\begin{figure}[!t]
\centering
\includegraphics[width=\columnwidth]{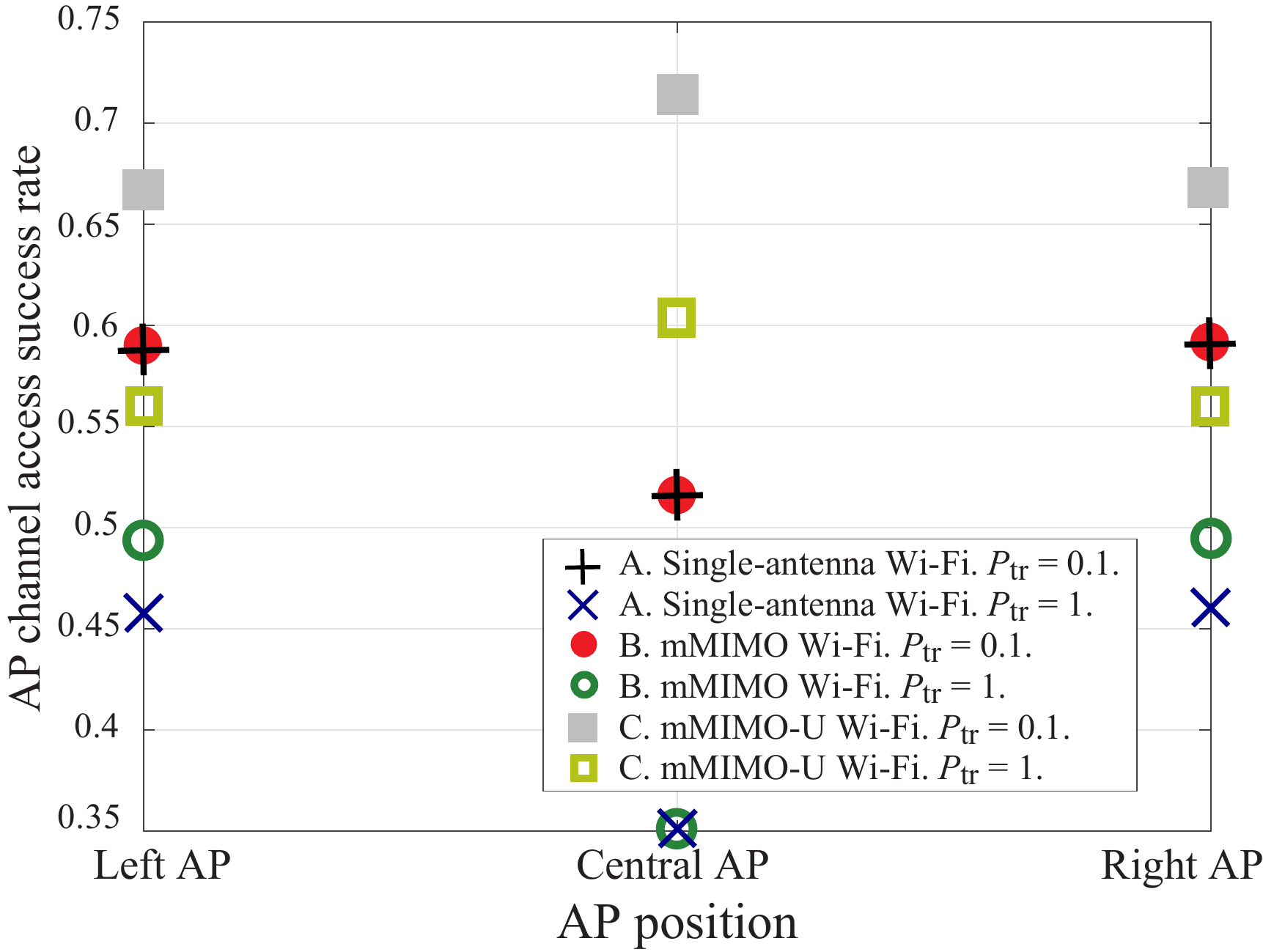}
\caption{Channel access success rate for the single-antenna Wi-Fi, mMIMO Wi-Fi, and mMIMO-U Wi-Fi scenarios with $P_{\text{tr}} = \left\lbrace 0.1, 1 \right\rbrace$.}
\label{fig:successfulChannelAccessProb}
\end{figure}

Fig.~\ref{fig:successfulChannelAccessProb} represents the rate of successful AP channel access for the three scenarios, also affected by the physical AP location. Intuitively, all success rates degrade for increasing traffic, a direct consequence of having to contend with more active nodes for channel access. By comparing scenarios A and B, i.e., single-antenna Wi-Fi and mMIMO Wi-Fi, we observe that the probability of successful channel access for the central AP does not change. This is because the same LBT procedure is adopted in both scenarios. Interestingly, the outermost APs slightly increase their access rate in scenario B, due to the mandatory power reduction at the central AP as per \eqref{eq:txPowermMIMO}.

Fig.~\ref{fig:successfulChannelAccessProb} also shows the increased channel access opportunities attained in scenario C, when the central AP implements mMIMO-U. The improvement is particularly noticeable under heavy traffic, i.e., for $P_{\text{tr}} = 1$, when the central AP increases its channel access success rate by $71\%$ (from $35\%$ to $60\%$). This is due to the radiation nulls placed during the proposed eLBT phase. Importantly, since the same radiation nulls are employed during transmission, mMIMO-U also enhances the channel access success rate for the other two APs.

\subsection{Downlink User SINR}

\begin{figure}[!t]
\centering
\includegraphics[width=0.965\columnwidth]{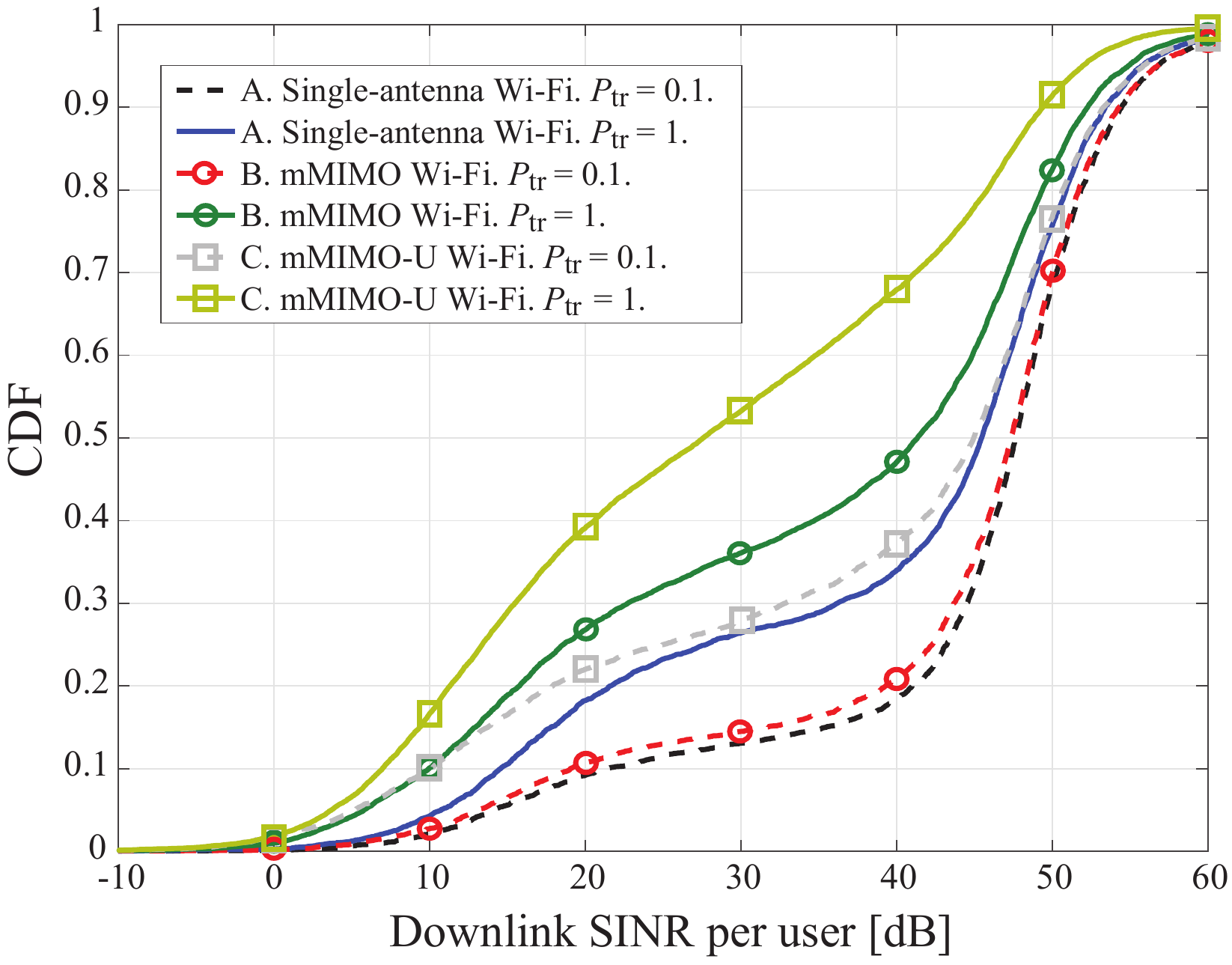} %
\caption{Downlink user SINR for the single-antenna Wi-Fi, mMIMO Wi-Fi, and mMIMO-U Wi-Fi scenarios with $P_{\text{tr}} = \left\lbrace 0.1, 1 \right\rbrace$.}
\label{fig:SINR}
\end{figure}

Fig.~\ref{fig:SINR} shows the cumulative distribution function (CDF) of the user SINR for the three scenarios following \eqref{eq:SINR}. The results in Fig.~\ref{fig:SINR} show a higher SINR under light traffic, because of a smaller number of interfering nodes. Scenario A achieves overall larger SINR values than scenarios B and C, again due to a reduced number of interfering STAs and APs. This is consistent with Fig.~\ref{fig:successfulChannelAccessProb}, which showed a lower AP channel access success rate for scenario A. The lowest SINR values in Fig.~\ref{fig:SINR} correspond to STAs that receive a strong UL-to-DL interference from a neighboring active STA. For scenario C, the $5\%$-worst SINR is bounded thanks to the proposed LBT/eLBT channel access selection with user scheduling, and amounts to $4$~dB. Paired with the larger number of spatially multiplexed users and the increased channel access opportunities provided by mMIMO-U, these SINRs yield a higher user throughput in scenario C, as shown in the following.
 
\vspace*{-0.3cm}
\subsection{Downlink Sum User Throughput}

\begin{figure}[!t]
\centering
\includegraphics[width=\columnwidth]{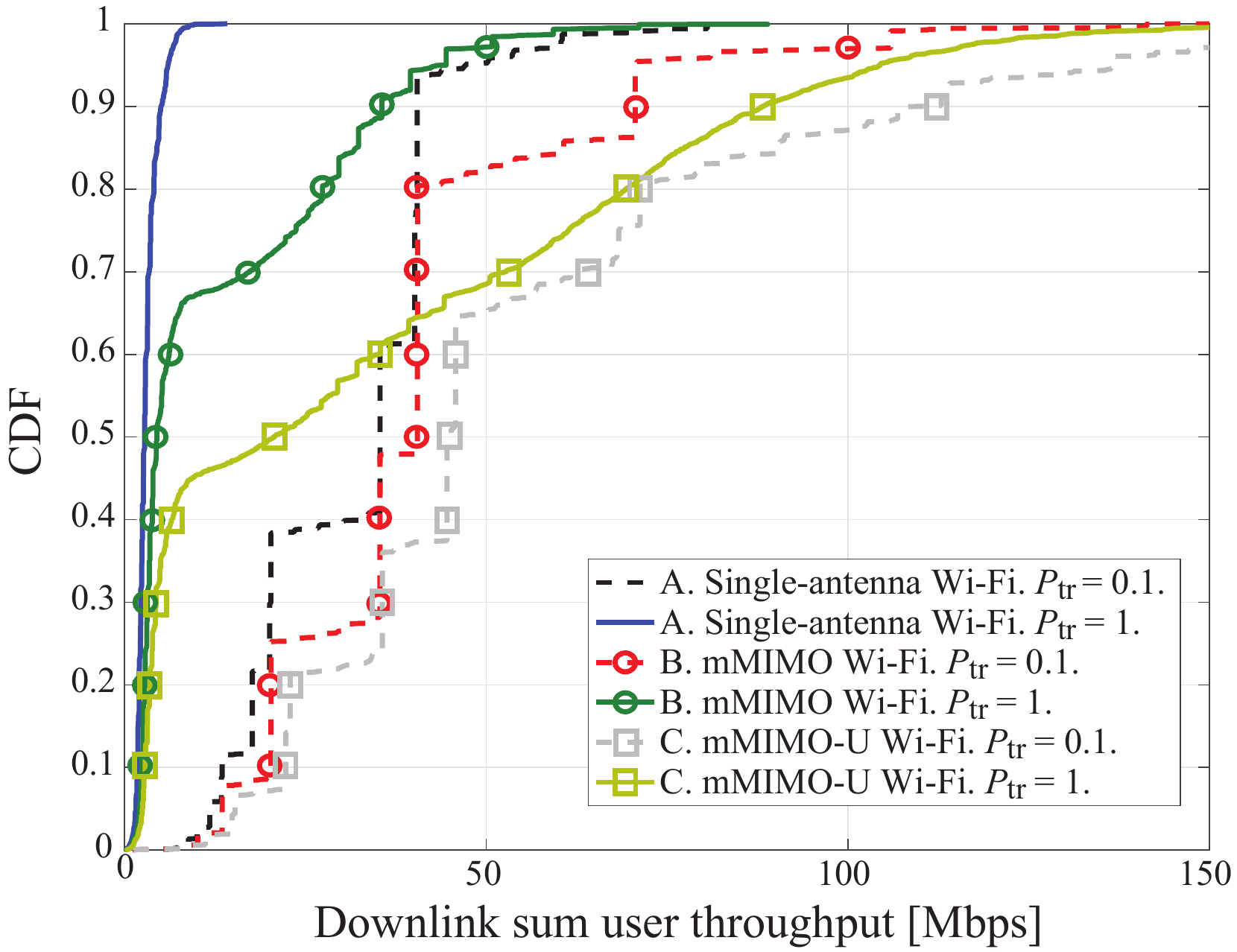}
\caption{Downlink sum user throughput for the single-antenna Wi-Fi, mMIMO Wi-Fi, and mMIMO-U Wi-Fi scenarios with $P_{\text{tr}} = \left\lbrace 0.1, 1 \right\rbrace$.}
\label{fig:downlinkSumUserThroughput}
\end{figure}

Fig.~\ref{fig:downlinkSumUserThroughput} shows the CDF of the DL sum user throughput, where the modulation and coding scheme is selected according to \cite{WLANStandard}. In this figure, we account for the fraction of time spent in UL/DL and for the service time per user. Note that the latter is affected by the channel access success rate and by the number of users performing time sharing. As a direct consequence, a user experiences significantly higher throughput under light traffic, i.e., for $P_{\text{tr}} = 0.1$. While all three scenarios achieve good throughput under light traffic, scenario C with mMIMO-U provides noticeable gains in the top $30$-th percentile.

The benefits of adopting mMIMO-U at the central AP are even more significant under heavy traffic, when a more frequent channel access facilitates serving each user more often. Moreover, the scheduling mechanism proposed in Section~III-C-3, which alternates LBT and eLBT CCA phases, ensures both (i) a guaranteed minimum throughput for users located near interfering nodes, and (ii) a very large sum throughput for users that are more immune to interference. Note that while the former are served after LBT and contribute to the lower part of the CDF, the latter are scheduled after eLBT and are captured in the upper part of the CDF. 
Overall, mMIMO-U demonstrates $4\times$ gains in the median throughput when compared to the mMIMO Wi-Fi scenario, and up to $7\times$ gains with respect to the single-antenna Wi-Fi setup.

\section{Conclusion}

We proposed and evaluated the performance of mMIMO-U in indoor deployments. We also devised a transmission pattern that schedules users and selects suitable CCA and interference suppression procedures depending on their resilience to interference. By alternating between more and less aggressive spatial reuse, our pattern ensures a guaranteed minimum service for users that are located near neighboring cells, as well as a very large throughput for users that are not. Our results demonstrate that mMIMO-U can significantly boost the sum user throughput when compared with conventional single- and multi-antenna solutions. Therefore, mMIMO-U can be regarded as a promising solution for future high-capacity indoor networks. 

We can identify multiple extensions of this work:

\begin{itemize}
\item \emph{Multi-channel operations:} Though we assumed that all devices share a single 20~MHz channel, this represents a worst-case scenario for mMIMO-U. In fact, it motivated the need of employing conventional LBT CCA 60\% of the time to serve users located near adjacent cells, thus capping the gains of mMIMO-U. Since multiple channels are available in the unlicensed band \cite{gast2005802}, future work should consider cross-channel user scheduling to mitigate inter-cell interference. This could provide further throughput gains thanks to a more frequent use of eLBT.

\item \emph{Multi-antenna users:} As shown in Fig.~\ref{fig:SINR}, scenarios with an increased number of simultaneous transmissions lead to a reduced SINR. In these setups, leveraging the presence of multiple antennas at the users could significantly enhance performance, e.g., through DL receive combining \cite{6509469}. In this way, both larger signal powers and reduced interference could be attained via additional beamforming gain and implicit interference rejection.

\item \emph{Multi-operator deployments:} We considered a single operator aiming to maximize the performance of its indoor network. Another scenario of interest is the one involving multiple operators, running their networks in the same area, and facing coexistence issues \cite{3GPP36889}. In this case, nodes that are physically close may belong to different closed subscriber groups, which conventionally forces them to alternate transmissions in an attempt to prevent collisions. Adopting mMIMO-U APs, operating on multiple channels and with multi-antenna users, could remove such need for time splitting, yielding a better spectrum reuse.
\end{itemize}


\ifCLASSOPTIONcaptionsoff
  \newpage
\fi
\bibliographystyle{IEEEtran}
\bibliography{Strings_Gio,Bib_Gio}
\end{document}